\documentclass[aps,twocolumn,showpacs,preprintnumbers,floats]{revtex4}\def\@cite#1#2{\textsuperscript{[{#1\if@tempswa , #2\fi}]}}

\usepackage{graphicx}
\usepackage{amsmath}
\usepackage{amsfonts}
\usepackage{amssymb}
\usepackage{color}
\usepackage{subfigure}
\usepackage{epsfig}
\usepackage{morefloats}
\usepackage{multirow}
\usepackage{graphicx,booktabs}
\usepackage{mathrsfs}
\usepackage{txfonts}
\usepackage{indentfirst}
\usepackage{graphicx,booktabs}

\usepackage{longtable,lscape}
\usepackage[colorlinks, citecolor=blue,anchorcolor=red,menucolor=red, linkcolor=red,filecolor=red,urlcolor=blue,frenchlinks=red]{hyperref}

\newcommand{\vsig}{\mbox{\boldmath$\sigma$\unboldmath}}

\usepackage{comment}

\begin{document}

%\begin{spacing}{2.0}

\title{Understanding the newly observed $\Xi_c^0$ states through their decays}

\author{Kai-Lei Wang$^{1}$~\footnote {Corresponding author, E-mail: wangkaileicz@foxmail.com}, Li-Ye Xiao$^{2,3}$~\footnote {Corresponding author, E-mail: lyxiao@ustb.edu.cn}, and Xian-Hui Zhong$^{3,4}$~\footnote {Corresponding author, E-mail: zhongxh@hunnu.edu.cn}}

\affiliation{ 1) Department
of Electronic Information and Physics, Changzhi University, Changzhi, Shanxi,046011,China}

\affiliation{ 2)School of Mathematics and Physics, University of Science and Technology Beijing,
Beijing 100083, China}

\affiliation{ 3) Department of
Physics, Hunan Normal University, and Key Laboratory of
Low-Dimensional Quantum Structures and Quantum Control of Ministry
of Education, Changsha 410081, China }
\affiliation{ 4) Synergetic
Innovation Center for Quantum Effects and Applications (SICQEA),
Hunan Normal University, Changsha 410081, China}

%\date{\today}

\begin{abstract}

Inspired by the newly observed $\Xi_c^0$ states by the LHCb Collaboration, we investigate the OZI-allowed two-body
strong decays of the $\lambda$-mode $1P$ wave $\Xi'_c$ states within the chiral quark model.
Our results indicate that: (i) the newly observed states $\Xi_c(2923)^0$ and $\Xi_c(2939)^0$ are good candidates of the $\lambda$-mode $1P$ wave $\Xi'_c$ states with the spin-parity $J^P=3/2^-$, namely $|^4P_{\lambda}3/2^-\rangle$ and $|^2P_{\lambda}3/2^-\rangle$, respectively. (ii) The another newly observed state $\Xi_c(2965)^0$ mostly corresponds to the $\lambda$-mode $1P$-wave $\Xi'_c$ state with the spin-parity $J^P=5/2^-$, namely $|^4P_{\lambda}5/2^-\rangle$. (iii) For the two $\lambda$-mode $J^P=1/2^-$ mixed states, the $|P_{\lambda}~1/2^-\rangle_1$ is a narrow state with a width of $\Gamma\sim15$ MeV and mainly decays into $\Xi'_c\pi$; while the $|P_{\lambda}~1/2^-\rangle_2$ state has a width of $\Gamma\sim52$ MeV and dominantly decays into $\Xi_c\pi$ and $\Lambda_cK$ channels. If the broad structure around $2880$ MeV observed at LHCb arises from the new $\Xi^0_c$ state, this state is very likely to be the $|P_{\lambda}~1/2^-\rangle_2$ state.

\end{abstract}

\pacs{}

\maketitle

\section{Introduction}

As an important kind of singly heavy baryons, the charmed-strange baryon $\Xi_c'$ spectrum
belonging to the flavor sextet $\mathbf{6}_F$ plays a crucial role in perfecting baryon spectra.
Although there are some discussions for the $\Xi_c'$ states in theory during the past several decades~\cite{Chen:2016spr,Cheng:2015iom,Crede:2013sze,Amhis:2019ckw}, no obvious progress of
the observations for the $\Xi_c'$ state has been achieved in experiments~\cite{Tanabashi:2018oca}.
In the $\Xi_c'$ spectrum, only the two ground states $\Xi_c'$ with $J^P=1/2^+$ and $\Xi_c^*(2645)$
with $J^P=3/2^+$ ($1S$ wave) have been established. So far, the low-lying $P$-wave $\Xi'_c$
states predicted in the quark model are still missing. It should be mentioned that in 2007
a structure $\Xi_c(2930)^0$ was observed by \emph{BaBar} in the $\Lambda^+_cK^-$ mass spectrum in
$B^-\rightarrow K^-\Lambda^+_c\bar{\Lambda}_c^-$ process~\cite{Aubert:2007eb}.
Later, the $\Xi_c(2930)^0$ was confirmed by the Belle Collaboration in the same decay process~\cite{Li:2017uvv},
while its charged partner $\Xi_c(2930)^+$ was also observed in $\bar{K}^{0} \Lambda _{c}^{+}$ final state
in the reaction $\bar{B}^{0} \rightarrow \bar{K}^{0} \Lambda _{c}^{+} \bar{\Lambda }_{c}^{-}$~\cite{Li:2018fmq}.
In addition, another structure $\Xi_c(2970)^0$ was first observed by \emph{BaBar}
in the $\Sigma_c(2455)^0K^0_S$ decay mode~\cite{Aubert:2007dt}. This structure was also
observed in both $\Xi'^{+}_c\pi^-$~\cite{Yelton:2016fqw} and $\Xi_c(2645)^{+}\pi^-$~\cite{Lesiak:2008wz} final states at Belle.
The structures $\Xi_c(2930)^0$ and $\Xi_c(2970)^0$ may be some signals of the missing
$P$-wave $\Xi'_c$ states first observed in experiments.

Very recently, the LHCb Collaboration observed three new states $\Xi_c(2923)^0$, $\Xi_c(2939)^0$, and $\Xi_c(2965)^0$  in
the $\Lambda_c^+K^-$ mass spectrum with a large significance~\cite{Aaij:2020yyt}, and their masses and natural widths
are determined to be
\begin{eqnarray}
 m[\Xi_c(2923)^0]& =&2923.04\pm0.59~\text{MeV},\nonumber\\
\Gamma[\Xi_c(2923)^0]& =&7.1\pm2.6~\text{MeV},\nonumber
\end{eqnarray}
\begin{eqnarray}
m[\Xi_c(2939)^0]& =&2938.55\pm0.52~\text{MeV},\nonumber\\
\Gamma[\Xi_c(2939)^0]& =&10.2\pm1.9~\text{MeV},\nonumber
\end{eqnarray}
\begin{eqnarray}
m[\Xi_c(2965)^0]& =&2964.88\pm0.54~\text{MeV},\nonumber\\
\Gamma[\Xi_c(2965)^0]& =&14.1\pm2.2~\text{MeV}.\nonumber
\end{eqnarray}
As pointed out in Ref.~\cite{Aaij:2020yyt}, the $\Xi_c(2930)^0$ observed in $B^-\rightarrow K^-\Lambda^+_c\bar{\Lambda}_c^-$ process~\cite{Aubert:2007eb,Li:2017uvv} might be due to the overlap of the two new narrower states, $\Xi_c(2923)^0$ and $\Xi_c(2939)^0$.
Stimulated by these three newly observed states, recently some groups have discussed their nature.
In Ref.~\cite{Yang:2020zjl},  these three newly observed states $\Xi_c(2923)^0$, $\Xi_c(2939)^0$, and $\Xi_c(2965)^0$
are suggested to be assigned as the $P$-wave $\Xi'_c$ states.  In Ref.~\cite{Lv:2020qi},  $\Xi_c(2923)^0$ and $\Xi_c(2939)^0$ may be good candidates of the P-wave states with $J^P=3/2^-$ and $5/2^-$ states and $\Xi_c(2965)^0$ can be assigned as $\Xi_c'(2S)$ state.
 To understand the nature of these newly observed states and
clarify whether they can be identified as the $\Xi'_c(1P)$ and $\Xi_c'(2S)$ state or not, more theoretical analysis is urgently needed.

%The newly observed $\Xi_c(2965)^0$ might be the same state as the known $\Xi_c(2970)^0$
%baryon~\cite{Aubert:2007dt,Yelton:2016fqw,Lesiak:2008wz}.

For the mass spectrum of the $\Xi_c'$ baryon, there exist many calculations with various models and effective theories in the literature~\cite{Chen:2016iyi,Roberts:2007ni,Ebert:2011kk,Ebert:2007nw,Faustov:2018vgl,Yamaguchi:2014era,Padmanath:2017lng,Zhang:2008pm,Wang:2010it,Chen:2015kpa,Yang:2020zrh}. We collect the predicted masses in Table~\ref{sp1}. From this table, it is found that the masses of the three $\Xi_c^0$ states observed by the LHCb Collaboration~\cite{Aaij:2020yyt} are in the predicted region of the $\lambda$-mode $1P$ wave $\Xi'_c$ excitations. Here, ``$\lambda$-mode" denotes one orbital excitation in a Jacobi coordinate between the light quarks and the heavy $c$ quark. Moreover, the possibility as $2S$ excitations cannot be excluded absolutely based simply on the predicted masses. In addition, it should be emphasized that in the light of the equal spacing rule~\cite{GellMann:1962xb,Okubo:1961jc}, the $\Xi_c(2923)^0$, $\Xi_c(2939)^0$ and $\Xi_c(2965)^0$ states probably correspond to their flavour multiplets $\Omega_c(3050)^0$, $\Omega_c(3065)^0$ and $\Omega_c(3090)^0$, respectively. Meanwhile, based on our previous work~\cite{Wang:2017kfr}, $\Omega_c(3050)^0$ and $\Omega_c(3065)^0$ could be assigned to be two $J^P=3/2^-$ states, $|1^4P_{\lambda}3/2^-\rangle$ and $|1^2P_{\lambda}3/2^-\rangle$, respectively; $\Omega_c(3090)^0$ very likely corresponds to the $J^P=5/2^-$ state $|1^4P_{\lambda}5/2^-\rangle$. To this extent, $\Xi_c(2923)^0$ and $\Xi_c(2939)^0$ are likely to have $J^P=3/2^-$, and $\Xi_c(2965)^0$ has $J^P=5/2^-$.
Besides mass spectra, the radiative and strong decay properties are crucial in pining down the inner structures of a state. Before the LHCb's measurement~\cite{Aaij:2020yyt}, there are also some discussions of the radiative and strong decay properties of the $1P$ wave $\Xi_c$ states~\cite{Chen:2007xf,Cheng:2006dk,Chen:2015kpa,Liu:2012sj,Wang:2017kfr,Cheng:2015naa,Aliev:2009jt,Chen:2017sci,Yang:2019cvw,Yang:2020zrh,Ye:2017yvl}.

In our previous works~\cite{Wang:2017kfr,Liu:2012sj}, the decay properties of the $1P$ wave $\Xi_c$ states were estimated with a chiral quark model, the mostly predicted decay widths of the $\lambda$-mode $1P$ wave $\Xi'_c$ states were about a few dozen MeV, which were roughly consistent with the LHCb's measurement~\cite{Aaij:2020yyt}. In the present work, by combining the newest data we further analyze the strong decay properties of the $\lambda$-mode $1P$ wave $\Xi'_c$ states with the chiral quark model, and attempt to put forward views on the inner structures of the three $\Xi_c^0$ states observed by the LHCb Collaboration~\cite{Aaij:2020yyt}.

This paper is organized as follows. In Sec. II we give a brief
introduction of the strong decay model. We discuss the strong decays of the low-lying $\lambda$-mode $1P$ wave $\Xi'_c$ states in Sec. III and summarize our results in
Sec. IV.

\begin{table*}[htp]
\begin{center}
\caption{\label{sp1}  The mass spectrum of $\Xi'_c$ belonging to $\mathbf{6}_F$ up to the $1P$-wave states in various models and effective theories. The $\Xi'_c$ states are denoted by $|N^{2S+1}L_{\sigma}J^P\rangle$ in the $LS$ coupling scheme. The unit of mass is MeV in the table.}
\begin{tabular}{lccccccccccccccccccccccccccccccccccccccccccccc}\hline\hline
%State                            ~~~~ &             ~~~~ &Predicted                ~~~~&Predicted  ~~~~&Predicted   ~~~~&Predicted  ~~~~&Predicted  ~~~~&Predicted ~~~~&        \\
$|N^{2S+1}L_{\sigma}J^P\rangle$  ~~~~&Ref.~\cite{Chen:2016iyi}  ~~~~&Ref.~\cite{Roberts:2007ni}  ~~~~&Ref.~\cite{Ebert:2011kk}   ~~~~&Ref.~\cite{Ebert:2007nw}  ~~~~&Ref.~\cite{Faustov:2018vgl}  ~~~~&Ref.~\cite{Yamaguchi:2014era} ~~~~&Observed state \\ \hline
$|0^2S \frac{1}{2}^+\rangle$                   ~~~~&2579~~~~&2592~~~~&2579~~~~&2578~~~~&2579~~~~&2594~~~~&$\Xi'^{+(0)}_c$ \\
$|1^4S \frac{3}{2}^+\rangle$                         ~~~~&2649~~~~&2650~~~~&2649~~~~&2654~~~~&2649~~~~&2649~~~~&$\Xi^*_c$  \\
$|1^2P_{\lambda} \frac{1}{2}^-\rangle$~~~~&2839~~~~&2859~~~~&2854~~~~&2928~~~~&2936~~~~&$\cdot\cdot\cdot$~~~~&  \\
$|1^2P_{\lambda} \frac{3}{2}^-\rangle$~~~~&2921~~~~&2871~~~~&2935~~~~&2931~~~~&$\cdot\cdot\cdot$~~~~&$\cdot\cdot\cdot$~~~~&$\Xi_c(2939)^0$ ?   \\
$|1^4P_{\lambda} \frac{1}{2}^-\rangle$~~~~&2900~~~~&$\cdot\cdot\cdot$~~~~&2936~~~~&2934~~~~&2935~~~~&$\cdot\cdot\cdot$& \\
$|1^4P_{\lambda} \frac{3}{2}^-\rangle$~~~~&2932   ~~~~&$\cdot\cdot\cdot$  ~~~~&2912    ~~~~&2900    ~~~~&$\cdot\cdot\cdot$~~~~&2866 ~~~~&$\Xi_c(2923)^0$ ? \\
$|1^4P_{\lambda} \frac{5}{2}^-\rangle$~~~~&2927   ~~~~&2905               ~~~~&2929    ~~~~&2921    ~~~~&2929   ~~~~ &2895~~~~ &$\Xi_c(2965)^0$ ?   \\
\hline\hline
\end{tabular}
\end{center}
\end{table*}

\section{ the model}

In this work we apply the chiral quark model~\cite{Manohar:1983md} to study the strong decay properties. Within the chiral quark model, the effective low energy quark-pseudoscalar-meson coupling in the SU(3) flavor basis at tree level is adopted as~\cite{Manohar:1983md}
\begin{eqnarray}\label{coup}
H_m=\sum_j
\frac{1}{f_m}\bar{\psi}_j\gamma^{j}_{\mu}\gamma^{j}_{5}\psi_j\vec{\tau}\cdot
\partial^{\mu}\vec{\phi}_m,
\end{eqnarray}
where $f_m$ is the pseudoscalar meson decay constant; $\psi_j$ represents the $j$th quark field in a baryon, and $\phi_m$ denotes the pseudoscalar meson octet.

To match the nonrelativistic harmonic oscillator spatial wave function in our calculations, one should adopt a nonrelativistic form of the quark-pseudoscalar coupling~\cite{Zhao:2002id,Li:1994cy,Li:1997gd}
\begin{eqnarray}\label{non-relativistic-expans}
H^{nr}_{m}&=&\sum_j\Big\{\frac{\omega_m}{E_f+M_f}\vsig_j\cdot
\textbf{P}_f+ \frac{\omega_m}{E_i+M_i}\vsig_j \cdot
\textbf{P}_i \nonumber\\
&&-\vsig_j \cdot \textbf{q} +\frac{\omega_m}{2\mu_q}\vsig_j\cdot
\textbf{p}'_j\Big\}I_j e^{-i\mathbf{q}\cdot \mathbf{r}_j},
\end{eqnarray}
where ($\omega_m,~\mathbf{q}$) denote the energy and three-vector momentum of the final light pseudoscalar meson; $(E_i,~M_i,~\mathbf{P}_i)$ and $(E_f,~M_f,~\mathbf{P}_f)$ are the energy, mass and three-vector momentum of the initial and final baryons, respectively. The $\mathbf{p}'_j (=\mathbf{p}_j-(m_j/M)\mathbf{P}_{\text{c.m.}})$ stands for the internal momentum of the $j$th quark in the baryon rest frame; $\mathbf{\sigma}_j$ sands for the Pauli spin vector on the $j$th quark; $\mu_q$ represents the reduced mass expressed as $1/\mu_q=1/m_j+1/m'_j$. The isospin operator $I_j$ associated with $\pi$ and $K$ mesons is given by
\begin{equation}
I_{j}=\begin{cases}
                             \frac{1}{\sqrt{2}}[a^{\dagger}_j(u)a_j(u)-a^{\dagger}_j(d)a_j(d)]     &$for$~\pi^{0},\\
                             a^{\dagger}_j(u)a_j(d)     &$for$~\pi^{-},\\
                             a^{\dagger}_j(u)a_j(s)     &$for$~K^{-}.

       \end{cases}
\end{equation}
Here, $a_j^{\dagger}(u,d)$ and $a_j(u,d,s)$ are the creation and annihilation operator for the $u,~d$ and $u,~d,~s$ quarks on $j$th quark, respectively.

Then the partial decay width for the emission of a light pseudoscalar meson in a hadron strong decay can be calculated with~\cite{Zhong:2007gp,Zhong:2008kd}
\begin{equation}\label{dww}
\Gamma=\left(\frac{\delta}{f_m}\right)^2\frac{(E_f +M_f)|\mathbf{q}|}{4\pi
M_i}\frac{1}{2J_i+1}\sum_{J_{iz}J_{fz}}|\mathcal{M}_{J_{iz},J_{fz}}|^2,
\end{equation}
where $\delta$ is a global parameter accounting for the strength of the quark-meson couplings; $J_{iz}$ and $J_{fz}$ are the third components of the total angular momenta of the initial and final baryons, respectively; $\mathcal{M}_{J_{iz},J_{fz}}$ denotes the transition amplitude.

With momentum $\mathbf{q}$ of the final light pseudoscalar meson increasing, the relativistic effect should be significant~\cite{sengl:2006ph}. To partly remedy the inadequacy of the nonrelativistic wave function as the momentum $\mathbf{q}$ increases, a commonly used Lorentz boost factor $\gamma_f$ is introduced into the decay amplitudes~\cite{Zhong:2008kd,Li:1995si,Zhao:1998fn,Zhong:2007fx}
\begin{eqnarray}
\mathcal{M}(\mathbf{q})\rightarrow \gamma_f\mathcal{M}(\gamma_f\mathbf{q}),
\end{eqnarray}
where $\gamma_f\equiv M_f/E_f$. In most decays, the corrections from the Lorentz boost are not drastic and the nonrelativistic prescription is reasonable.

The model parameters have been well determined in previous works~\cite{Wang:2017kfr,Liu:2012sj}, and we collect them in Table~\ref{masses}.
In the calculations the masses of the final baryons and mesons are
taken from the PDG~\cite{Tanabashi:2018oca} and collected in Table~\ref{masses} as well.
The harmonic oscillator space-wave functions $\Psi^n_{lm}=R_{nl}Y_{lm}$ are adopted to describe the spatial wave
function of the initial and final baryons, and the harmonic oscillator parameter $\alpha_{\rho}$
in the wave functions for $ds/us$ system is taken as $\alpha_{\rho}=420$ MeV.
Another harmonic oscillator parameter $\alpha_{\lambda}$ can be related to $\alpha_{\rho}$
with the relation  $\alpha_{\lambda}=[3m_c/(2m_q+m_c)]^{1/4}\alpha_{\rho}$, where $m_q$ denotes
the light quark mass.

\begin{table}[h]
\caption{\label{masses} The parameters and final hadron' masses~\cite{Tanabashi:2018oca} used
in this work. The unit is MeV except the parameter $\delta$, which is a dimensionless quantity. $\Xi^*_c$ denotes $\Xi_c(2645)$ in the table.}
\begin{tabular}{llllccccccccccccccl}\hline\hline
State~~~~~~~&Mass ~~~~~~~&~~~~State~~~~~~&Mass\\
$\Xi_c^+$~~~~~~~&2467.9~~~~~~~&~~~~$\pi^-$~~~~~~&139.6\\
$\Xi_c^0$~~~~~~~&2470.9~~~~~~~&~~~~$\pi^0$~~~~~~&135.0\\
$\Xi_c'^+$~~~~~~~&2577.4~~~~~~~&~~~~$K^-$~~~~~~&493.8\\
$\Xi_c'^0$~~~~~~~&2578.8~~~~~~~&~~~~$\Lambda_c^+$~~~~~~&2286.5\\
$\Xi_c^{*+}$~~~~~~~&2645.5~~~~~~~&~~~~$\Xi_c^{*0}$~~~~~~&2646.3\\
\hline
\multicolumn{2}{c}{Parameter}~~~~&$\delta$~~~~~~&0.557\\
\multicolumn{2}{c}{}        ~~~~&$f_\pi$~~~~~~&132\\
\multicolumn{2}{c}{}         ~~~~&$f_K$~~~~~~&160 \\
\multicolumn{2}{c}{}         ~~~~&$\alpha_{\rho}$~~~~~~& 420 \\
\hline
\multicolumn{2}{c}{Constituent quark mass}&$m_q$~~~~~~~~&330\\
\multicolumn{2}{c}{}                      &$m_s$~~~~~~~~&450\\
 \multicolumn{2}{c}{}                     &$m_c$~~~~~~~~&1480\\
\hline\hline
\end{tabular}
\end{table}

\section{Results and analysis}

The masses of the three $\Xi_c^0$ states newly observed by the LHCb Collaboration~\cite{Aaij:2020yyt} are in the predicted mass region of the $\lambda$-mode $1P$ wave $\Xi'_c$ states(see Table~\ref{sp1}). To clarify the possibility and further investigate their inner structures, we conduct a systematic study of the strong decay properties for the $\lambda$-mode $1P$ wave $\Xi'_c$ states within the framework of a chiral quark model. Our results and theoretical predictions are presented as follows.

%\begin{figure}[]
%\centering \epsfxsize=6.0cm \epsfbox{pwave.eps} \caption{The
%strong decay properties of the $1P$-wave states in the $N=1$ shell. }\label{fig-pwave}
%\end{figure}

\subsection{$1P$ states with $J^P=1/2^-$}

\begin{figure}[ht]
\centering\epsfxsize=7 cm \epsfbox{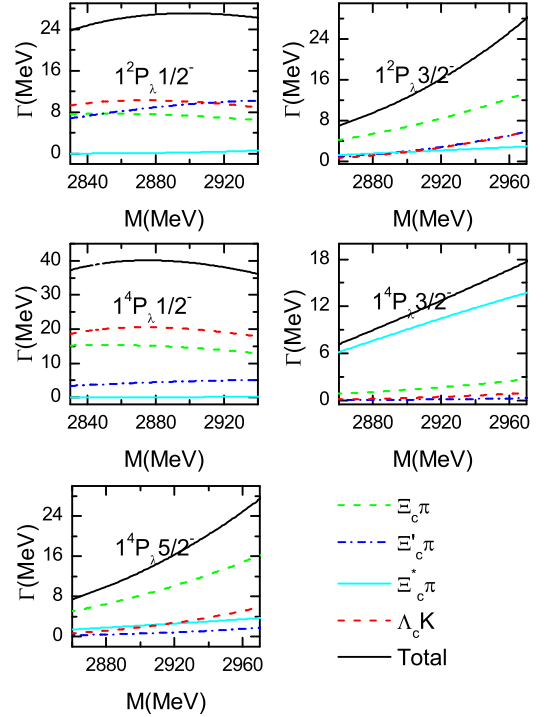} \vspace{-0.4 cm} \caption{The strong decay partial width for the $1P$ wave $\Xi'_c$ states as a function of mass.}\label{p-mass}
\end{figure}

In the $\Xi'_c$ family, there are two $\lambda$-mode $J^P=1/2^-$ states $|1^2P_{\lambda}1/2^-\rangle$ and $|1^4P_{\lambda}1/2^-\rangle$.
Their masses are predicted to be about $M=(2830-2940)$ MeV (see Table~\ref{sp1}), and their OZI-allowed two-body strong decay channels are $\Xi_c\pi$, $\Xi'_c\pi$, $\Xi^*_c\pi$ and $\Lambda_c^+K^-$. Considering the uncertainties of the predicted masses, we plot the partial decay widths as functions of the masses of the states $\Xi'_c|1^2P_{\lambda}1/2^-\rangle$ and $\Xi'_c|1^4P_{\lambda}1/2^-\rangle$ in Fig.~\ref{p-mass}. Their decay properties remain relatively stable within the mass range of $(2830-2940)$ MeV.

For the state $\Xi'_c|1^2P_{\lambda}1/2^-\rangle$, the total decay width varies in the region of $\Gamma\sim(23-27)$ MeV, and the dominant decay modes are $\Xi_c\pi$, $\Xi'_c\pi$ and $\Lambda_cK$. The branching fraction for each of the dominant decay mode is about 30\%. For the other $J^P=1/2^-$ state $\Xi'_c|1^4P_{\lambda}1/2^-\rangle$, the total decay width is about $\Gamma\sim(29-40)$ MeV with mass varied in the region what we considered. It mainly decays into $\Xi_c\pi$ and $\Lambda_cK$. Meanwhile, the partial decay width of the $\Xi'_c\pi$ mode is significant with a branching ratio of $\sim$10\%.

\begin{figure}[ht]
\centering \epsfxsize=5 cm \epsfbox{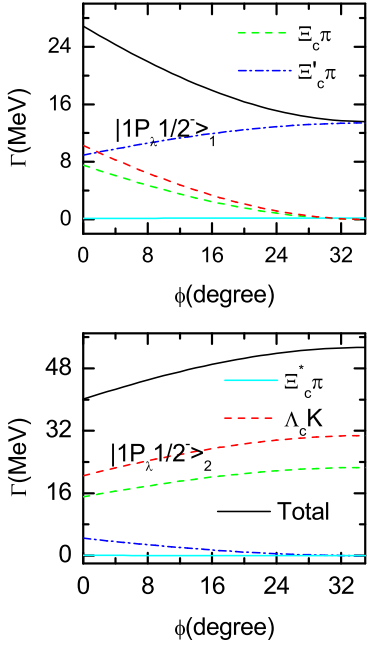} \vspace{-0.5 cm} \caption{The strong decay partial width for the $J^P=1/2^-$ mixed $\Xi'_c$ states as a function of the mixing angle $\phi$. The masses of the mixed states $|1P_{\lambda}~1/2^-\rangle_1$ and $|1P_{\lambda}~1/2^-\rangle_2$ are fixed at $M=2880$ MeV. }\label{p-angle}
\end{figure}

We notice that for the singly heavy flavour quark systems, proper consideration of the heavy quark symmetry is necessary~\cite{Wang:2018fjm}. Namely, the physical states with the spin-parity $J^P=1/2^-$ are very likely to be mixed states between $|1^2P_{\lambda}1/2^-\rangle$ and $|1^4P_{\lambda}1/2^-\rangle$ by the following mixing scheme,
\begin{equation}\label{mixp}
\left(\begin{array}{c}| 1P_{\lambda}~{1/2^-}\rangle_1\cr |  1P_{\lambda}~{1/2^-}\rangle_2
\end{array}\right)=\left(\begin{array}{cc} \cos\phi & \sin\phi \cr -\sin\phi &\cos\phi
\end{array}\right)
\left(\begin{array}{c} |1^2P_{\lambda}1/2^-
\rangle \cr |1^4P_{\lambda}1/2^-\rangle
\end{array}\right).
\end{equation}
The mixing angle $\phi$ may rang from $\phi=0^\circ$ to that of heavy quark symmetry limit ($\phi=35^\circ$). In our previous work~\cite{Wang:2017kfr,Wang:2017hej}, we obtained that the state $\Omega_c(3000)$ could be explained as the mixed state $|1P_{\lambda}~1/2^-\rangle_1$ with a mixing angle $\phi\simeq24^\circ$. As the same flavour multiplet, the mixing angle in the $\Xi'_c$ family should be comparable with that in the $\Omega_c$ family. Meanwhile, according to the equal spacing rule~\cite{GellMann:1962xb,Okubo:1961jc}, the mass of the mixed state $|1P_{\lambda}~1/2^-\rangle_1$ in the $\Xi'_c$ family is lighter about $\sim$120 MeV than that of state in the $\Omega_c$ family, namely $M_{\Xi'_c|1P_{\lambda}~1/2^-\rangle_1}\simeq2880$ MeV. Thus according to the mixing scheme defined in Eq.~(\ref{mixp}), in Fig.~\ref{p-angle} we plot the strong decay width of the $|1P_{\lambda}~1/2^-\rangle_1$ state as a function of the mixing angle $\phi$ in the range of ($0^{\circ}$, $35^{\circ}$) by fixing the mass at $M=2880$ MeV.

It is found that with the mixing angle increasing, the partial decay widths of the $\Lambda_cK$ and $\Xi_c\pi$ modes for $\Xi'_c|1P_{\lambda}~1/2^-\rangle_1$ are rapidly suppressed and the $\Xi'_c\pi$ decay channel almost saturates its total decay width.
Taking a possible mixing angle $\phi\simeq 24^\circ$ constrained by the $\Omega_c(3000)$ state,
we find the $\Xi'_c|1P_{\lambda}~1/2^-\rangle_1$ state has a narrow width of $\Gamma\sim 15.0$ MeV,
and the branching fraction of the dominant decay channel $\Xi'_c\pi$ is
\begin{eqnarray}
\mathcal{B}[\Xi'_c|1P_{\lambda}~1/2^-\rangle_1\rightarrow \Xi'_c\pi]\simeq 85\%.
\end{eqnarray}
The decay rate of $\Xi'_c|1P_{\lambda}~\frac{1}{2}^-\rangle_1$ into $\Lambda_cK$
strongly depends on the mixing angle. If taking a slightly larger mixing
$\phi\simeq 28^\circ$ than that ($\phi\simeq 24^\circ$) determined by
$\Omega_c(3000)$, one finds the decay rate into the $\Lambda_cK$ channel is nearly zero.
Thus, the $\Xi'_c|1P_{\lambda}~\frac{1}{2}^-\rangle_1$ may be hardly observed
in the $\Lambda_cK$ final state.

Furthermore, the predicted decay width of the mixed state $\Xi'_c|1P_{\lambda}~1/2^-\rangle_1$ seems to be comparable
with the decay width of the three $\Xi^0_c$ states observed by the LHCb Collaboration~\cite{Aaij:2020yyt}. However, it should be kept in mind that the mass of $\Xi'_c|1P_{\lambda}~1/2^-\rangle_1$ should be around $\sim2880$ MeV considering a reasonable explanation for the properties of $\Omega_c(3000)$. Thus, most likely the three $\Xi^0_c$ states as the state $\Xi'_c|1P_{\lambda}~1/2^-\rangle_1$ is excluded. Considering the predicted width of $\Xi'_c|1P_{\lambda}~1/2^-\rangle_1$ being narrow, this state might be observed in the $\Xi'_c\pi$ channel when enough data are accumulated in experiments.

\begin{table*}[htp]
\begin{center}
\caption{\label{decaywidth}  The decay properties of the $P$-wave $\Xi'_c$ states compared with the  observations. $\Gamma^{\text{th}}_{\text{total}}$ presents the total decay width calculated in the present work, while $\Gamma^{\text{exp}}_{\text{total}}$ presents the total width obtained from the LHCb experiment~\cite{Aaij:2020yyt}. The units of mass and width are MeV in the table.}
\begin{tabular}{cccccccccccccccccccccccccccccccccccccccccccccc}\hline\hline
State~~~~~~ &Mass  ~~~~~~&$\Gamma[\Xi_c\pi]$  ~~~~~~&$\Gamma[\Xi'_c\pi]$   ~~~~~~&$\Gamma[\Xi^*_c\pi]$  ~~~~~~&$\Gamma[\Lambda_cK]$  ~~~~~~&$\Gamma^{\text{th}}_{\text{total}}$ ~~~~~~&$\Gamma^{\text{exp}}_{\text{total}}$ ~~~~~~&Possible assignment       \\
 \hline
$|1P_{\lambda}~\frac{1}{2}^-\rangle_1$~~~~~~&2880~~~~~~& 0.86~~~~~~&12.9~~~~~~&0.18~~~~~~&1.17~~~~~~&15.1~~~~~~&$\cdot\cdot\cdot$~~~~~~&$\cdot\cdot\cdot$ \\
$|1P_{\lambda}~\frac{1}{2}^-\rangle_2$~~~~~~&2880~~~~~~& 21.7~~~~~~&0.51~~~~~~&0.01~~~~~~&29.6~~~~~~&51.8~~~~~~&$\cdot\cdot\cdot$~~~~~~&$\cdot\cdot\cdot$ \\
$|1^4P_{\lambda}~\frac{3}{2}^-\rangle$~~~~~~&2923~~~~~~& 1.74~~~~~~&0.15~~~~~~&10.7~~~~~~&0.48~~~~~~&13.1~~~~~~&$7.1\pm0.8\pm1.8$~~~~~~&$\Xi_c(2923)^0$\\
$|1^2P_{\lambda}~\frac{3}{2}^-\rangle$~~~~~~&2939~~~~~~& 10.2~~~~~~&3.80~~~~~~&2.46~~~~~~&3.74~~~~~~&20.2~~~~~~&$10.2\pm0.8\pm1.1$~~~~~~&$\Xi_c(2939)^0$ \\
$|1^4P_{\lambda}~\frac{5}{2}^-\rangle$~~~~~~&2965~~~~~~& 15.5~~~~~~&1.64~~~~~~&3.57 ~~~~~~&5.43~~~~~~&26.1~~~~~~&$14.1\pm0.9\pm1.3$~~~~~~&$\Xi_c(2965)^0$ \\
\hline\hline
\end{tabular}
\end{center}
\end{table*}

The other mixed state $|1P_{\lambda}~1/2^-\rangle_2$ should be a relatively broad state
with a width much larger than the three newly observed $\Xi_c^0$ states. At this moment we investigate its width range with $M$=2880 MeV in Fig.~\ref{p-angle} as well. From the figure, the total decay width is about $\Gamma\sim(38-50)$ MeV with the mixing angle varying in range of ($0^{\circ}$, $35^{\circ}$). Taking a possible mixing angle $\phi=24^{\circ}$ constrained by the $\Omega_c(3000)$ state,
we find that the mixed state $\Xi'_c|1P_{\lambda}~1/2^-\rangle_2$ has a width of $\Gamma\sim52$ MeV, and dominantly decays
into the $\Xi_c\pi$ and $\Lambda_cK$ final states with comparable branching fractions
\begin{eqnarray}
\mathcal{B}[\Xi'_c|1P_{\lambda}~1/2^-\rangle_2\rightarrow \Xi_c\pi]\simeq 42\%,\\
\mathcal{B}[\Xi'_c|1P_{\lambda}~1/2^-\rangle_2\rightarrow \Lambda_cK]\simeq 57\%.
\end{eqnarray}

%\begin{eqnarray}
%\frac{\Gamma[|1P_{\lambda}~1/2^-\rangle_2\rightarrow \Xi_c\pi]}{\Gamma[|1P_{\lambda}~1/2^-\rangle_2\rightarrow \Lambda_cK]}\sim0.84.
%\end{eqnarray}
%The predicted branching ratios are

Both the mass and decay width of this mixed state are inconsistent with the observations of the three newly observed $\Xi_c^0$ states, thus, the three $\Xi^0_c$ states as the state $\Xi'_c|1P_{\lambda}~1/2^-\rangle_2$ should be excluded. Since $\Xi'_c|1P_{\lambda}~1/2^-\rangle_2$ is not a very broad state, it might be observed in the $\Xi_c\pi$ and $\Lambda_cK$ channels.

It should be remarked that according to the LHCb' measurement~\cite{Aaij:2020yyt}, the $\Lambda_cK$ mass spectrum shows
a broad structure around $2880$ MeV, which might be due to the presence of additional new $\Xi^0_c$ states. Combining the predicted decay properties of $\Xi'_c|1P_{\lambda}~1/2^-\rangle_2$ in this work, if the broad structure in the region around $M\sim2880$ MeV arises from a new $\Xi^0_c$ state, this state is very likely to be the $\Xi'_c|1P_{\lambda}~1/2^-\rangle_2$ state. Moreover, the possibility of the broad structure arising from the overlapping of $\Xi'_c|1P_{\lambda}~1/2^-\rangle_1$ and $\Xi'_c|1P_{\lambda}~1/2^-\rangle_2$ cannot be ruled out as well.

\subsection{$1P$ states with $J^P=3/2^-$}

There are two $\lambda$-mode $J^P=3/2^-$ states $\Xi'_c|1^2P_{\lambda}3/2^-\rangle$ and $\Xi'_c|1^4P_{\lambda}3/2^-\rangle$. The predicted masses of these two states are listed in Table~\ref{sp1}. From the table, it is seen that predicted masses are about $\sim2930$ MeV, which are comparable with the masses of $\Xi_c^0(2923)$ and $\Xi_c^0(2939)$ measured by the LHCb Collaboration~\cite{Aaij:2020yyt}. As the good candidates of $\Xi_c^0(2923)$ and $\Xi_c^0(2939)$, it is essential to study the decay properties of the $\Xi'_c|1^2P_{\lambda}3/2^-\rangle$ and $\Xi'_c|1^4P_{\lambda}3/2^-\rangle$.

We plot the decay properties of the $\Xi'_c|1^2P_{\lambda}3/2^-\rangle$ and $\Xi'_c|1^4P_{\lambda}3/2^-\rangle$
as functions of their masses in the range of $M=(2860-2970)$ MeV in Fig.~\ref{p-mass} as well. It is found that the total decay width of $\Xi'_c|1^2P_{\lambda}3/2^-\rangle$ is about $\Gamma\sim(7-28)$ MeV and that of $\Xi'_c|1^4P_{\lambda}3/2^-\rangle$ is $\Gamma\sim(7-18)$ MeV with masses increasing in the range what we considered. The dominant decay mode of $\Xi'_c|1^2P_{\lambda}3/2^-\rangle$ is $\Xi_c\pi$, while $\Xi'_c|1^4P_{\lambda}3/2^-\rangle$ mainly decays into $\Xi'^{*}_c\pi$ channel.

The predicted decay widths of $\Xi'_c|1^2P_{\lambda}3/2^-\rangle$ and $\Xi'_c|1^4P_{\lambda}3/2^-\rangle$ are also comparable with the observed values of $\Xi_c^0(2939)$ and $\Xi_c^0(2923)$ within the uncertainties. Meanwhile, with the similar masses, $|1^2P_{\lambda}3/2^-\rangle$ is slightly broader than $|1^4P_{\lambda}3/2^-\rangle$. By combining the width order $\Gamma[\Xi_c(2939)^0]>\Gamma[\Xi_c(2923)^0]$ from experiments and our theoretical predictions, we may conclude that $\Xi_c(2939)^0$ and $\Xi_c(2923)^0$ prefer to the $\Xi'_c|1^2P_{\lambda}3/2^-\rangle$ and $\Xi'_c|1^4P_{\lambda}3/2^-\rangle$ states, respectively.
In addition, according to the equal spacing rule~\cite{GellMann:1962xb,Okubo:1961jc}, the $\Xi_c(2923)^0$ and $\Xi_c(2939)^0$ states probably correspond to their flavour multiplets $\Omega_c(3050)^0$ and $\Omega_c(3065)^0$, respectively. In our previous work~\cite{Wang:2017kfr,Wang:2017hej}, $\Omega_c(3050)^0$ could be assigned to be  $|1^4P_{\lambda}3/2^-\rangle$ and $\Omega_c(3065)^0$ was assigned to be $|1^2P_{\lambda}3/2^-\rangle$.
To this extend, the $\Xi_c(2923)^0$ and $\Xi_c(2939)^0$ states assigned to be $|1^4P_{\lambda}3/2^-\rangle$ and $|1^2P_{\lambda}3/2^-\rangle$, respectively, is reasonable as well.

Fixing the mass of $\Xi'_c|1^4P_{\lambda}3/2^-\rangle$ with $M=2923$ MeV, we obtain
\begin{eqnarray}
\Gamma_{\text{total}}[\Xi'_c|1^4P_{\lambda}3/2^-\rangle]\simeq13~\text{MeV}.
\end{eqnarray}
The predicted branching fraction of the dominant decay mode $\Xi^{*}_c\pi$ is
\begin{eqnarray}
\mathcal{B}[\Xi'_c|1^4P_{\lambda}3/2^-\rangle\rightarrow \Xi^{*}_c\pi]\simeq 82\%.
\end{eqnarray}
Meanwhile, the decay rates of $\Xi'_c|1^4P_{\lambda}3/2^-\rangle$ into $\Xi_c\pi$ and $\Lambda_cK$
are considerable, and the predicted branching fractions are
\begin{eqnarray}
\mathcal{B}[\Xi'_c|1^4P_{\lambda}3/2^-\rangle\rightarrow \Xi_c\pi]\simeq 13\%,\\
\mathcal{B}[\Xi'_c|1^4P_{\lambda}3/2^-\rangle\rightarrow \Lambda_cK]\simeq 4\%.
\end{eqnarray}
The sizeable branching fraction for $\Xi'_c|1^4P_{\lambda}3/2^-\rangle$ into $\Lambda_cK$
is consistent with the nature of $\Xi_c(2923)^0$, which is observed in the $\Lambda_cK$ channel. If $\Xi_c(2923)^0$ corresponds to the state $\Xi'_c|1^4P_{\lambda}3/2^-\rangle$ indeed, the $\Xi_c\pi$ and $\Xi^{*}_c\pi$ also may be measured in future experiments due to their large branching fractions.

In the same way, we fix the mass of $\Xi'_c|1^2P_{\lambda}3/2^-\rangle$ with $M=2939$ MeV.
Then, the total decay width is predicted to be
\begin{eqnarray}
\Gamma_{\text{total}}[\Xi'_c|1^2P_{\lambda}3/2^-\rangle]\simeq20~\text{MeV}.
\end{eqnarray}
This state mainly decays into $\Xi_c\pi$ channel with the branching fraction
\begin{eqnarray}
\mathcal{B}[\Xi'_c|1^2P_{\lambda}3/2^-\rangle\rightarrow \Xi_c\pi]\simeq 51\%.
\end{eqnarray}
The partial decay widths of the other three decay channels $\Xi'_c\pi$, $\Xi^{*}_c\pi$, and $\Lambda_cK$ are comparable. The partial decay ratios are
\begin{eqnarray}
\frac{\Gamma[\Xi'_c|1^2P_{\lambda}3/2^-\rangle\rightarrow \Lambda_cK]}{\Gamma[\Xi'_c|1^2P_{\lambda}3/2^-\rangle\rightarrow \Xi'_c\pi]}\simeq0.98,\\
\frac{\Gamma[\Xi'_c|1^2P_{\lambda}3/2^-\rangle\rightarrow \Xi^*_c\pi]}{\Gamma[\Xi'_c|1^2P_{\lambda}3/2^-\rangle\rightarrow \Xi'_c\pi]}\simeq0.65.
\end{eqnarray}
If $\Xi_c(2939)^0$ observed in the $\Lambda_cK$ channel corresponds to the state $\Xi'_c|1^2P_{\lambda}3/2^-\rangle$ indeed, the comparable partial decay widths indicate this state may be established in the $\Xi_c\pi$, $\Xi'_c\pi$ and $\Xi^*_c\pi$ decay channels as well in future experiments.

Moreover, assigning the $\Xi_c(2923)^0$ and $\Xi_c(2939)^0$ to $\Xi'_c|1^4P_{\lambda}3/2^-\rangle$ and $\Xi'_c|1^2P_{\lambda}3/2^-\rangle$, respectively, we notice that the predicted total decay width ratio
\begin{eqnarray}
R_1=\frac{\Gamma_{\text{total}}[\Xi'_c|1^4P_{\lambda}3/2^-\rangle]}{\Gamma_{\text{total}}[\Xi'_c|1^2P_{\lambda}3/2^-\rangle]}\simeq0.65,
\end{eqnarray}
highly agrees with the observed central value $R_1^{\text{exp}}=\frac{\Gamma_{\text{total}}[\Xi_c(2923)^0]}{\Gamma_{\text{total}}[\Xi_c(2939)^0]}\simeq0.70$.

\subsection{$1P$ states with $J^P=5/2^-$}

There is only one $\lambda$-mode $J^P=5/2^-$ state $|1^4P_{\lambda}5/2^-\rangle$. The predicted mass of this state is listed in Table~\ref{sp1}. In terms of the predicted mass, the possibility of the three newly observed $\Xi^0_c$ states taken as the state $\Xi'_c|1^4P_{\lambda}5/2^-\rangle$ cannot be excluded. To investigate the effects of the uncertainties of the mass on the decay properties of the $\Xi'_c|1^4P_{\lambda}5/2^-\rangle$ state, we show the variation of the partial decay width with the change of mass in Fig.~\ref{p-mass}. From the figure, the variation curves between the partial decay width and the mass for this state is similar to that for $\Xi'_c|1^2P_{\lambda}3/2^-\rangle$. The dominant decay mode for $\Xi'_c|1^4P_{\lambda}5/2^-\rangle$ is $\Xi_c\pi$. Meanwhile, the partial decay width of the $\Lambda_cK$ mode is sizeable, and becomes more and more significant with the mass increasing in the region of $(2860-2970)$ MeV.

The predicted decay properties of $\Xi'_c|1^4P_{\lambda}5/2^-\rangle$ are consistent with the observations of $\Xi_c(2965)^0$. Meanwhile, considering the equal spacing rule~\cite{GellMann:1962xb,Okubo:1961jc}, $\Xi_c(2965)^0$ most likely corresponds to its flavour mulitiplet $\Omega_c(3090)$, which was assigned to the $|1^4P_{\lambda}5/2^-\rangle$ state according to our previous study~\cite{Wang:2017kfr}. Thus, it is reasonable to assign $\Xi_c(2965)^0$ as the $\Xi'_c|1^4P_{\lambda}5/2^-\rangle$ state.

According to our calculations, we get
\begin{eqnarray}
\Gamma_{\text{total}}[\Xi'_c|1^4P_{\lambda}5/2^-\rangle]\simeq26.1~\text{MeV}
\end{eqnarray}
with a mass of $M=2965$ MeV (see Table~\ref{decaywidth}). The dominant decay mode is $\Xi_c\pi$ with the branching fraction
\begin{eqnarray}
\mathcal{B}[\Xi'_c|1^4P_{\lambda}5/2^-\rangle\rightarrow \Xi_c\pi]\simeq 59\%.
\end{eqnarray}
The decay rate of $\Xi'_c|1^4P_{\lambda}5/2^-\rangle$ into the $\Lambda_cK$ channel is significant as well, and the predicted branching ratio is
\begin{eqnarray}
\mathcal{B}[\Xi'_c|1^4P_{\lambda}5/2^-\rangle\rightarrow \Lambda_cK]\simeq 21\%.
\end{eqnarray}
The sizeable branching fraction of $\Xi'_c|1^4P_{\lambda}5/2^-\rangle$ into $\Lambda_cK$ is consistent with the observation of the $\Xi_c(2965)^0$ signal in the $\Lambda_cK$ decay channel.

In addition, the total decay width ratios among the $\Xi'_c|1^4P_{\lambda}3/2^-\rangle$, $\Xi'_c|1^2P_{\lambda}3/2^-\rangle$, and $\Xi'_c|1^4P_{\lambda}5/2^-\rangle$ states are predicted to be
\begin{eqnarray}
R_2=\frac{\Gamma_{\text{total}}[\Xi'_c|1^4P_{\lambda}3/2^-\rangle]}{\Gamma_{\text{total}}[\Xi'_c|1^4P_{\lambda}5/2^-\rangle]}\simeq0.50,\\
R_3=\frac{\Gamma_{\text{total}}[\Xi'_c|1^2P_{\lambda}3/2^-\rangle]}{\Gamma_{\text{total}}[\Xi'_c|1^4P_{\lambda}5/2^-\rangle]}\simeq0.77.
\end{eqnarray}
The predicted ratios are good consistent with the experimental central values ($R_2^{\text{exp}}=\frac{\Gamma_{\text{total}}[\Xi_c(2923)^0]}{\Gamma_{\text{total}}[\Xi_c(2965)^0]}\simeq0.50$ and $R_3^{\text{exp}}=\frac{\Gamma_{\text{total}}[\Xi_c(2939)^0]}{\Gamma_{\text{total}}[\Xi_c(2965)^0]}\simeq0.77$)
of those among the three states $\Xi_c(2923)^0$, $\Xi_c(2939)^0$, and $\Xi_c(2965)^0$, respectively.

Finally, it should be mentioned that the known $\Xi_c(2970)^0$ was observed
in the $\Sigma_c(2455)^0K^0_S$~\cite{Aubert:2007dt}, $\Xi'^{+}_c\pi^-$~\cite{Yelton:2016fqw},
and $\Xi_c(2645)^{+}\pi^-$~\cite{Lesiak:2008wz,Yelton:2016fqw} decay modes may correspond 
two different resonances with a very similar mass. The $\Xi_c(2970)^0$ was observed
in the $\Sigma_c(2455)^0K^0_S$~\cite{Aubert:2007dt} final state cannot be
considered as the same state of $\Xi_c(2965)^0$ although they have a very similar mass,
because the $\Sigma_c(2455)^0K^0_S$ mode of $\Xi_c(2965)^0$ is forbidden as the
$\Xi'_c|1^4P_{\lambda}3/2^-\rangle$ state. However, one cannot exclude the $\Xi_c(2965)^0$
resonance as the same resonance observed in the $\Xi'^{+}_c\pi^-$~\cite{Yelton:2016fqw},
and $\Xi_c(2645)^{+}\pi^-$~\cite{Lesiak:2008wz,Yelton:2016fqw} final states. 
The $\Xi_c(2970)$ observed in the $\Sigma_c(2455)^0K^0_S$ final state may be explained with the $\rho$-mode $1P$ wave 
$\Xi'_c$ states~\cite{Liu:2012sj}, or the first positive parity excitations 
of the $\Xi_c$~\cite{Cheng:2015naa,Chen:2007xf}.

\section{Summary}

In this paper, we carry out a systematic study of the OZI allowed two-body strong decays of the $\lambda$-mode $1P$ wave $\Xi'_c$ states in the framework of a chiral quark model. Combining our theoretical predictions and the experimental observations,
we give possible interpretations for the three new states $\Xi_c(2923)^0$, $\Xi_c(2939)^0$, and $\Xi_c(2965)^0$ observed by the LHCb Collaboration.

Our theoretical results show that the newly observed states $\Xi_c(2923)^0$ and $\Xi_c(2939)^0$ are most likely to be explained as the $\lambda$-mode $1P$ wave $\Xi'_c$ states with spin-parity $J^P=3/2^-$, namely $\Xi'_c|1^4P_{\lambda}3/2^-\rangle$ and $\Xi'_c|1^2P_{\lambda}3/2^-\rangle$, respectively. The $\Xi_c(2923)^0$ and $\Xi_c(2939)^0$ may be flavour partners of $\Omega_c(3050)^0$ and $\Omega_c(3065)^0$, respectively. Meanwhile, if the arrangements in this work are correct, then the dominant decay mode of $\Xi_c(2923)^0$ is $\Xi'^*_c\pi$ and that of $\Xi_c(2939)^0$ is $\Xi_c\pi$. This can be tested in future experiments.

The another newly observed state $\Xi_c(2965)^0$ may corresponds to the $\lambda$-mode $1P$-wave $\Xi'_c$ state with spin-parity $J^P=5/2^-$, namely $\Xi'_c|1^4P_{\lambda}5/2^-\rangle$. The $\Xi_c(2965)^0$ may be a flavour partner of $\Omega_c(3090)^0$. Besides the $\Lambda_cK$ decay channel, the decay rate of $\Xi'_c|1^4P_{\lambda}5/2^-\rangle$ into $\Xi_c\pi$ is significant as well, and the predicted branching ratio is about 59\%. The large branching fraction indicates that this state may be reconstructed in the $\Xi_c\pi$ decay channel as well.

There are strong configuration mixings in the $J^P=1/2^-$ $\lambda$-mode states.
The mixed state $\Xi'_c|1P_{\lambda}~1/2^-\rangle_1$ might be a flavour partner of
$\Omega_c(3000)^0$. This $J^P=1/2^-$ mixed state may have a mass of $M\sim2880$ MeV and a narrow width
of $\Gamma\sim15$ MeV. The dominant decay mode of $|1P_{\lambda}~1/2^-\rangle_1$ is $\Xi'_c\pi$ with a
branching fraction of $> 85\%$. The decay rate into $\Lambda_cK$ is strongly suppressed due to
the heavy quark symmetry. The $\Xi'_c|1P_{\lambda}~1/2^-\rangle_1$ may be observed in
the $\Xi'_c\pi$ final state.

The other $J^P=1/2^-$ mixed state $\Xi'_c|1P_{\lambda}~1/2^-\rangle_2$ has
a relatively broad width of $\Gamma\sim48$ MeV, which is about 3 times larger
than that of $\Xi'_c|1P_{\lambda}~1/2^-\rangle_1$. The $\Xi'_c|1P_{\lambda}~1/2^-\rangle_2$
mainly decays into $\Xi_c\pi$ and $\Lambda_cK$ channels with branching ratios 45\% and 54\%, respectively.
Considering the mass and decay properties, if the broad structure in the $\Lambda_cK$ mass spectrum around $M\sim2880$ MeV observed by the LHCb
Collaboration arises from a new $\Xi^0_c$ state, this state is very likely to be the $\Xi'_c|1P_{\lambda}~1/2^-\rangle_2$ state. Moreover, the possibility of the broad structure arising from the overlapping of $\Xi'_c|1P_{\lambda}~1/2^-\rangle_1$ and $\Xi'_c|1P_{\lambda}~1/2^-\rangle_2$ cannot be ruled out as well.

Finally it should be mentioned that combining our previous study~\cite{Xiao:2020oif} of the newly observed $\Omega_b^*$ states at LHCb~\cite{Aaij:2020cex},
we find that the $\Xi_c(2923)^0$, $\Xi_c(2939)^0$, and $\Xi_c(2965)^0$ may be flavour partners of $\Omega_b(6330)$,
$\Omega_b(6340)$, and $\Omega_b(6350)$, respectively. The missing mixed state $\Xi'_c|1P_{\lambda}~1/2^-\rangle_1$ may be
a flavour partner of $\Omega_b(6316)$.

\section*{Acknowledgements }
This work is supported by the National Natural
Science Foundation of China under Grants No. 11775078, No. 11947048 and  No. U1832173.

%\end{spacing}


\begin{thebibliography}{99}

%\cite{Chen:2016spr}
\bibitem{Chen:2016spr}
  H.~X.~Chen, W.~Chen, X.~Liu, Y.~R.~Liu and S.~L.~Zhu,
  A review of the open charm and open bottom systems,
  Rept. Prog. Phys. 80, 076201 (2017).
  %doi:10.1088/1361-6633/aa6420
  %[arXiv:1609.08928 [hep-ph]].
  %%CITATION = doi:10.1088/1361-6633/aa6420;%%
  %160 citations counted in INSPIRE as of 05 Apr 2020


%\cite{Crede:2013sze}
\bibitem{Crede:2013sze}
  V.~Crede and W.~Roberts,
  Progress towards understanding baryon resonances,
  Rept. Prog. Phys.  76, 076301 (2013).
  %doi:10.1088/0034-4885/76/7/076301
  %[arXiv:1302.7299 [nucl-ex]].
  %%CITATION = doi:10.1088/0034-4885/76/7/076301;%%
  %165 citations counted in INSPIRE as of 05 Apr 2020


%\cite{Amhis:2019ckw}
\bibitem{Amhis:2019ckw}
  Y.~S.~Amhis {\it et al.} [HFLAV Collaboration],
  Averages of $b$-hadron, $c$-hadron, and $\tau$-lepton properties as of 2018,
  arXiv:1909.12524 [hep-ex].
  %%CITATION = ARXIV:1909.12524;%%
  %60 citations counted in INSPIRE as of 05 Apr 2020


%\cite{Cheng:2015iom}
\bibitem{Cheng:2015iom}
  H.~Y.~Cheng,
  Charmed baryons circa 2015,
  Front.\ Phys.\ (Beijing) {\bf 10}, 101406 (2015).
  %doi:10.1007/s11467-015-0483-z
  %%CITATION = doi:10.1007/s11467-015-0483-z;%%
  %41 citations counted in INSPIRE as of 08 Jan 2020


%\cite{Tanabashi:2018oca}
\bibitem{Tanabashi:2018oca}
  M.~Tanabashi {\it et al.} [Particle Data Group],
  Review of Particle Physics,
  Phys.\ Rev.\ D {\bf 98}, 030001 (2018).
  %doi:10.1103/PhysRevD.98.030001
  %%CITATION = doi:10.1103/PhysRevD.98.030001;%%
  %3466 citations counted in INSPIRE as of 06 Jan 2020


%\cite{Aubert:2007eb}
\bibitem{Aubert:2007eb}
  B.~Aubert {\it et al.} [BaBar Collaboration],
  A Study of $\bar{B}\rightarrow \Xi_c \bar{\Lambda}_c$ and $\bar{B}\rightarrow \Lambda^+_c \bar{\Lambda}^+_c \bar{K}$ decays at BABAR,
  Phys. Rev. D 77, 031101 (2008).
  %doi:10.1103/PhysRevD.77.031101
 % [arXiv:0710.5775 [hep-ex]].
  %%CITATION = doi:10.1103/PhysRevD.77.031101;%%
  %47 citations counted in INSPIRE as of 03 Apr 2020

%\cite{Li:2017uvv}
\bibitem{Li:2017uvv}
  Y.~B.~Li {\it et al.} [Belle Collaboration],
  Observation of $\Xi_{c}(2930)^0$ and updated measurement of $B^{-} \to K^{-} \Lambda_{c}^{+} \bar{\Lambda}_{c}^{-}$ at Belle, 
  Eur. Phys. J. C  78, 252 (2018).
  %doi:10.1140/epjc/s10052-018-5720-5
  %[arXiv:1712.03612 [hep-ex]].
  %%CITATION = doi:10.1140/epjc/s10052-018-5720-5;%%
  %25 citations counted in INSPIRE as of 03 Apr 2020


%\cite{Li:2018fmq}
\bibitem{Li:2018fmq}
  Y.~B.~Li {\it et al.} [Belle Collaboration],
  Evidence of a structure in $\bar{K}^{0} \Lambda _{c}^{+}$ consistent with a charged $\Xi _c(2930)^{+}$ , and updated measurement of $\bar{B}^{0} \to \bar{K}^{0} \Lambda _{c}^{+} \bar{\Lambda }_{c}^{-}$ at Belle,
  Eur. Phys. J. C  78, 928 (2018).
  %doi:10.1140/epjc/s10052-018-6425-5
  %[arXiv:1806.09182 [hep-ex]].
  %%CITATION = doi:10.1140/epjc/s10052-018-6425-5;%%
  %14 citations counted in INSPIRE as of 05 Apr 2020


%\cite{Aubert:2007dt}
\bibitem{Aubert:2007dt}
  B.~Aubert {\it et al.} [BaBar Collaboration],
  A Study of Excited Charm-Strange Baryons with Evidence for new Baryons $\Xi_c(3055)^+$ and $\Xi_c(3123)^+$, 
  Phys. Rev. D 77, 012002 (2008).
  %doi:10.1103/PhysRevD.77.012002
  %[arXiv:0710.5763 [hep-ex]].
  %%CITATION = doi:10.1103/PhysRevD.77.012002;%%
  %93 citations counted in INSPIRE as of 03 Apr 2020



%\cite{Yelton:2016fqw}
\bibitem{Yelton:2016fqw}
  J.~Yelton {\it et al.} [Belle Collaboration],
  Study of Excited $\Xi_c$ States Decaying into $\Xi_c^0$ and $\Xi_c^+$ Baryons,
  Phys. Rev. D  94, 052011 (2016).
 % doi:10.1103/PhysRevD.94.052011
 % [arXiv:1607.07123 [hep-ex]].
  %%CITATION = doi:10.1103/PhysRevD.94.052011;%%
  %39 citations counted in INSPIRE as of 03 Apr 2020


%\cite{Lesiak:2008wz}
\bibitem{Lesiak:2008wz}
  T.~Lesiak {\it et al.} [Belle Collaboration],
  Measurement of masses of the $\Xi_c(2645)$ and $\Xi_c(2815)$ baryons and observation of $\Xi_c(2980) \rightarrow \Xi_c(2645)\pi$,
  Phys. Lett. B  665, 9 (2008).
  %doi:10.1016/j.physletb.2008.05.055
  %[arXiv:0802.3968 [hep-ex]].
  %%CITATION = doi:10.1016/j.physletb.2008.05.055;%%
  %35 citations counted in INSPIRE as of 03 Apr 2020

%\cite{Aaij:2020yyt}
\bibitem{Aaij:2020yyt}
  R.~Aaij {\it et al.} [LHCb Collaboration],
  Observation of new $\Xi_c^0$ baryons decaying to $\Lambda_c^+ K^-$,
  arXiv:2003.13649 [hep-ex].
  %%CITATION = ARXIV:2003.13649;%%
  %1 citations counted in INSPIRE as of 03 Apr 2020

%\cite{Yang:2020zjl}
\bibitem{Yang:2020zjl}
  H.~M.~Yang, H.~X.~Chen and Q.~Mao,
  Identifying the $\Xi_c^0$ baryons observed by LHCb as $P$-wave $\Xi_c^\prime$ baryons,
  arXiv:2004.00531 [hep-ph].
  %%CITATION = ARXIV:2004.00531;%%

\bibitem{Lv:2020qi}
  Q.~F.~L\"{u},
  Canonical interpretations of the newly observed $\Xi_c(2923)^0$, $\Xi_c(2939)^0$ and $\Xi_c(2965)^0$ resonances,
  arXiv:2004.02374 [hep-ph].



%\cite{Chen:2016iyi}
\bibitem{Chen:2016iyi}
  B.~Chen, K.~W.~Wei, X.~Liu and T.~Matsuki,
  Low-lying charmed and charmed-strange baryon states,
  Eur. Phys. J. C  77, 154 (2017).
%  doi:10.1140/epjc/s10052-017-4708-x
%  [arXiv:1609.07967 [hep-ph]].
  %%CITATION = doi:10.1140/epjc/s10052-017-4708-x;%%
  %32 citations counted in INSPIRE as of 01 Apr 2020


%\cite{Faustov:2018vgl}
\bibitem{Faustov:2018vgl}
  R.~N.~Faustov and V.~O.~Galkin,
  Heavy baryon spectroscopy,
  EPJ Web Conf. 204, 08001 (2019).
  %doi:10.1051/epjconf/201920408001
  %[arXiv:1811.02232 [hep-ph]].
  %%CITATION = doi:10.1051/epjconf/201920408001;%%
  %2 citations counted in INSPIRE as of 03 Apr 2020


%\cite{Yamaguchi:2014era}
\bibitem{Yamaguchi:2014era}
  Y.~Yamaguchi, S.~Ohkoda, A.~Hosaka, T.~Hyodo and S.~Yasui,
 Heavy quark symmetry in multihadron systems,
  Phys. Rev. D  91, 034034 (2015).
 % doi:10.1103/PhysRevD.91.034034
 % [arXiv:1402.5222 [hep-ph]].
  %%CITATION = doi:10.1103/PhysRevD.91.034034;%%
  %34 citations counted in INSPIRE as of 03 Apr 2020



%\cite{Padmanath:2017lng}
\bibitem{Padmanath:2017lng}
  M.~Padmanath and N.~Mathur,
  Quantum Numbers of Recently Discovered $\Omega^{0}_{c}$ Baryons from Lattice QCD,
  Phys. Rev. Lett.  119, 042001 (2017).
  %doi:10.1103/PhysRevLett.119.042001
  %[arXiv:1704.00259 [hep-ph]].
  %%CITATION = doi:10.1103/PhysRevLett.119.042001;%%
  %51 citations counted in INSPIRE as of 05 Apr 2020

%\cite{Zhang:2008pm}
\bibitem{Zhang:2008pm}
  J.~R.~Zhang and M.~Q.~Huang,
  Heavy baryon spectroscopy in QCD,
  Phys. Rev. D 78, 094015 (2008).
  %doi:10.1103/PhysRevD.78.094015
 % [arXiv:0811.3266 [hep-ph]].
  %%CITATION = doi:10.1103/PhysRevD.78.094015;%%
  %60 citations counted in INSPIRE as of 05 Apr 2020


%\cite{Wang:2010it}
\bibitem{Wang:2010it}
  Z.~G.~Wang,
  Analysis of the ${1/2^-}$ and ${3/2^-}$ heavy and doubly heavy baryon states with QCD sum rules,
  Eur. Phys. J. A  47, 81 (2011).
  %doi:10.1140/epja/i2011-11081-8
  %[arXiv:1003.2838 [hep-ph]].
  %%CITATION = doi:10.1140/epja/i2011-11081-8;%%
  %48 citations counted in INSPIRE as of 05 Apr 2020

%%\cite{Chen:2015kpa}
%\bibitem{Chen:2015kpa}
%  H.~X.~Chen, W.~Chen, Q.~Mao, A.~Hosaka, X.~Liu and S.~L.~Zhu,
%  $P$-wave charmed baryons from QCD sum rules,
%  Phys. Rev. D  91, 054034 (2015).
%  %doi:10.1103/PhysRevD.91.054034
%  %[arXiv:1502.01103 [hep-ph]].
%  %%CITATION = doi:10.1103/PhysRevD.91.054034;%%
%  %66 citations counted in INSPIRE as of 05 Apr 2020


%\cite{Yang:2020zrh}
\bibitem{Yang:2020zrh}
  H.~M.~Yang and H.~X.~Chen,
  $P$-wave bottom baryons of the $SU(3)$ flavor $\mathbf{6}_F$,
  arXiv:2003.07488 [hep-ph].
  %%CITATION = ARXIV:2003.07488;%%
  %1 citations counted in INSPIRE as of 05 Apr 2020


%\cite{Ebert:2011kk}
\bibitem{Ebert:2011kk}
  D.~Ebert, R.~N.~Faustov, and V.~O.~Galkin,
  Spectroscopy and Regge trajectories of heavy baryons in the relativistic quark-diquark picture,
  Phys. Rev. D  84, 014025 (2011).
%  doi:10.1103/PhysRevD.84.014025
%  [arXiv:1105.0583 [hep-ph]].
  %%CITATION = doi:10.1103/PhysRevD.84.014025;%%
  %124 citations counted in INSPIRE as of 06 Aug 2019

%\cite{Roberts:2007ni}
\bibitem{Roberts:2007ni}
  W.~Roberts and M.~Pervin,
  Heavy baryons in a quark model,
  Int. J. Mod. Phys. A  23, 2817 (2008).
%  doi:10.1142/S0217751X08041219
%  [arXiv:0711.2492 [nucl-th]].
  %%CITATION = doi:10.1142/S0217751X08041219;%%
  %264 citations counted in INSPIRE as of 06 Aug 2019

%\cite{Ebert:2007nw}
\bibitem{Ebert:2007nw}
  D.~Ebert, R.~N.~Faustov and V.~O.~Galkin,
  Masses of excited heavy baryons in the relativistic quark model,
  Phys. Lett. B  659, 612 (2008).
%  doi:10.1016/j.physletb.2007.11.037
%  [arXiv:0705.2957 [hep-ph]].
  %%CITATION = doi:10.1016/j.physletb.2007.11.037;%%
  %181 citations counted in INSPIRE as of 03 Apr 2020

%\cite{Chen:2015kpa}
\bibitem{Chen:2015kpa}
  H.~X.~Chen, W.~Chen, Q.~Mao, A.~Hosaka, X.~Liu and S.~L.~Zhu,
  P-wave charmed baryons from QCD sum rules,
  Phys. Rev. D 91, 054034 (2015).
  %doi:10.1103/PhysRevD.91.054034
  %[arXiv:1502.01103 [hep-ph]].
  %%CITATION = doi:10.1103/PhysRevD.91.054034;%%
  %66 citations counted in INSPIRE as of 03 Apr 2020


%\cite{GellMann:1962xb}
\bibitem{GellMann:1962xb}
  M.~Gell-Mann,
  Symmetries of baryons and mesons,
  Phys. Rev. 125, 1067 (1962).
  %doi:10.1103/PhysRev.125.1067
  %%CITATION = doi:10.1103/PhysRev.125.1067;%%
  %1601 citations counted in INSPIRE as of 03 Apr 2020

%\cite{Okubo:1961jc}
\bibitem{Okubo:1961jc}
  S.~Okubo,
  Note on unitary symmetry in strong interactions,
  Prog. Theor. Phys.  27, 949 (1962).
  %doi:10.1143/PTP.27.949
  %%CITATION = doi:10.1143/PTP.27.949;%%
  %649 citations counted in INSPIRE as of 03 Apr 2020


%\cite{Wang:2017kfr}
\bibitem{Wang:2017kfr}
  K.~L.~Wang, Y.~X.~Yao, X.~H.~Zhong and Q.~Zhao,
  Strong and radiative decays of the low-lying $S$- and $P$-wave singly heavy baryons,
  Phys. Rev. D  96, 116016 (2017).
  %doi:10.1103/PhysRevD.96.116016
  %[arXiv:1709.04268 [hep-ph]].
  %%CITATION = doi:10.1103/PhysRevD.96.116016;%%
  %44 citations counted in INSPIRE as of 03 Apr 2020


%\cite{Chen:2007xf}
\bibitem{Chen:2007xf}
  C.~Chen, X.~L.~Chen, X.~Liu, W.~Z.~Deng and S.~L.~Zhu,
  Strong decays of charmed baryons,
  Phys. Rev. D 75, 094017 (2007).
  %doi:10.1103/PhysRevD.75.094017
  %[arXiv:0704.0075 [hep-ph]].
  %%CITATION = doi:10.1103/PhysRevD.75.094017;%%
  %102 citations counted in INSPIRE as of 03 Apr 2020

%\cite{Cheng:2006dk}
\bibitem{Cheng:2006dk}
  H.~Y.~Cheng and C.~K.~Chua,
  Strong Decays of Charmed Baryons in Heavy Hadron Chiral Perturbation Theory,
  Phys. Rev. D  75, 014006 (2007).
 % doi:10.1103/PhysRevD.75.014006
 % [hep-ph/0610283].
  %%CITATION = doi:10.1103/PhysRevD.75.014006;%%
  %108 citations counted in INSPIRE as of 03 Apr 2020

%\cite{Cheng:2015naa}
\bibitem{Cheng:2015naa}
  H.~Y.~Cheng and C.~K.~Chua,
  Strong Decays of Charmed Baryons in Heavy Hadron Chiral Perturbation Theory: An Update,
  Phys. Rev. D  92, 074014 (2015).
 % doi:10.1103/PhysRevD.92.074014
  %[arXiv:1508.05653 [hep-ph]].
  %%CITATION = doi:10.1103/PhysRevD.92.074014;%%
  %46 citations counted in INSPIRE as of 05 Apr 2020

%\cite{Aliev:2009jt}
\bibitem{Aliev:2009jt}
  T.~M.~Aliev, K.~Azizi and A.~Ozpineci,
  Radiative Decays of the Heavy Flavored Baryons in Light Cone QCD Sum Rules,
  Phys. Rev. D  79, 056005 (2009).
  %doi:10.1103/PhysRevD.79.056005
  %[arXiv:0901.0076 [hep-ph]].
  %%CITATION = doi:10.1103/PhysRevD.79.056005;%%
  %48 citations counted in INSPIRE as of 05 Apr 2020

%\cite{Chen:2017sci}
\bibitem{Chen:2017sci}
  H.~X.~Chen, Q.~Mao, W.~Chen, A.~Hosaka, X.~Liu and S.~L.~Zhu,
  Decay properties of $P$-wave charmed baryons from light-cone QCD sum rules,
  Phys. Rev. D  95, 094008 (2017).
  %doi:10.1103/PhysRevD.95.094008
  %[arXiv:1703.07703 [hep-ph]].
  %%CITATION = doi:10.1103/PhysRevD.95.094008;%%
  %58 citations counted in INSPIRE as of 05 Apr 2020

%\cite{Yang:2019cvw}
\bibitem{Yang:2019cvw}
  H.~M.~Yang, H.~X.~Chen, E.~L.~Cui, A.~Hosaka and Q.~Mao,
  Decay properties of P-wave bottom baryons within light-cone sum rules,
  Eur. Phys. J. C  80, 80 (2020).
  %doi:10.1140/epjc/s10052-020-7637-z
  %[arXiv:1909.13575 [hep-ph]].
  %%CITATION = doi:10.1140/epjc/s10052-020-7637-z;%%
  %6 citations counted in INSPIRE as of 05 Apr 2020


  %\cite{Liu:2012sj}
\bibitem{Liu:2012sj}
  L.~H.~Liu, L.~Y.~Xiao and X.~H.~Zhong,
  Charm-strange baryon strong decays in a chiral quark model,
  Phys. Rev. D  86, 034024 (2012).
  %doi:10.1103/PhysRevD.86.034024
  %[arXiv:1205.2943 [hep-ph]].
  %%CITATION = doi:10.1103/PhysRevD.86.034024;%%
  %35 citations counted in INSPIRE as of 03 Apr 2020


%\cite{Ye:2017yvl}
\bibitem{Ye:2017yvl}
  D.~D.~Ye, Z.~Zhao and A.~Zhang,
  Study of $P$-wave excitations of observed charmed strange baryons,
  Phys.\ Rev.\ D {\bf 96}, 114009 (2017).
 % doi:10.1103/PhysRevD.96.114009
 % [arXiv:1709.00689 [hep-ph]].
  %%CITATION = doi:10.1103/PhysRevD.96.114009;%%
  %16 citations counted in INSPIRE as of 18 Apr 2020

%\cite{Manohar:1983md}
\bibitem{Manohar:1983md}
  A.~Manohar and H.~Georgi,
  Chiral Quarks and the Nonrelativistic Quark Model,
  Nucl. Phys. B  234, 189 (1984).
  %doi:10.1016/0550-3213(84)90231-1
  %%CITATION = doi:10.1016/0550-3213(84)90231-1;%%
  %1931 citations counted in INSPIRE as of 26 May 2018

%\cite{Zhao:2002id}
\bibitem{Zhao:2002id}
  Q.~Zhao, J.~S.~Al-Khalili, Z.~P.~Li and R.~L.~Workman,
  Pion photoproduction on the nucleon in the quark model,
  Phys. Rev. C  65, 065204 (2002).
  %doi:10.1103/PhysRevC.65.065204
 % [nucl-th/0202067].
  %%CITATION = doi:10.1103/PhysRevC.65.065204;%%
  %52 citations counted in INSPIRE as of 13 Jan 2020

%\cite{Li:1994cy}
\bibitem{Li:1994cy}
  Z.~P.~Li,
  The Threshold pion photoproduction of nucleons in the chiral quark model,
  Phys. Rev. D  50, 5639 (1994).
  %doi:10.1103/PhysRevD.50.5639
  %[hep-ph/9404269].
  %%CITATION = doi:10.1103/PhysRevD.50.5639;%%
  %37 citations counted in INSPIRE as of 26 Jul 2017

%\cite{Li:1997gd}
\bibitem{Li:1997gd}
  Z.~P.~Li, H.~X.~Ye and M.~H.~Lu,
  An Unified approach to pseudoscalar meson photoproductions off nucleons in the quark model,
  Phys. Rev. C  56, 1099 (1997).
  %doi:10.1103/PhysRevC.56.1099
  %[nucl-th/9706010].
  %%CITATION = doi:10.1103/PhysRevC.56.1099;%%
  %73 citations counted in INSPIRE as of 26 Jul 2017


%\cite{Zhong:2007gp}{Zhong:2008kd}
\bibitem{Zhong:2007gp}
  X.~H.~Zhong and Q.~Zhao,
  Charmed baryon strong decays in a chiral quark model,
  Phys. Rev. D  77, 074008 (2008).
%  doi:10.1103/PhysRevD.77.074008
%  [arXiv:0711.4645 [hep-ph]].
  %%CITATION = doi:10.1103/PhysRevD.77.074008;%%
  %57 citations counted in INSPIRE as of 28 May 2018

%\cite{Zhong:2008kd}
\bibitem{Zhong:2008kd}
  X.~H.~Zhong and Q.~Zhao,
  Strong decays of heavy-light mesons in a chiral quark model,
  Phys. Rev. D  78, 014029 (2008).
  %doi:10.1103/PhysRevD.78.014029
  %[arXiv:0803.2102 [hep-ph]].
  %%CITATION = doi:10.1103/PhysRevD.78.014029;%%
  %75 citations counted in INSPIRE as of 26 May 2018

%\cite{sengl:2006ph}
\bibitem{sengl:2006ph}
 B.~Sengl, Ph.~D. thesis, University of Graz, Graz, 2006.


%\cite{Li:1995si}
\bibitem{Li:1995si}
  Z.~P.~Li,
  The Kaon photoproduction of nucleons in the chiral quark model,
  Phys. Rev. C  52, 1648 (1995).
  %doi:10.1103/PhysRevC.52.1648
  %[hep-ph/9502218].
  %%CITATION = doi:10.1103/PhysRevC.52.1648;%%
  %81 citations counted in INSPIRE as of 04 Apr 2020

%\cite{Zhao:1998fn}
\bibitem{Zhao:1998fn}
  Q.~Zhao, Z.~p.~Li and C.~Bennhold,
  Vector meson photoproduction with an effective Lagrangian in the quark model,
  Phys. Rev. C  58, 2393 (1998).
  %doi:10.1103/PhysRevC.58.2393
  %[nucl-th/9806100].
  %%CITATION = doi:10.1103/PhysRevC.58.2393;%%
  %95 citations counted in INSPIRE as of 04 Apr 2020


 %\cite{Zhong:2007fx}
\bibitem{Zhong:2007fx}
  X.~H.~Zhong, Q.~Zhao, J.~He and B.~Saghai,
  Study of $\pi^-p \rightarrow \eta n$ at low energies in a chiral constituent quark model,
  Phys. Rev. C 76, 065205 (2007).
  %doi:10.1103/PhysRevC.76.065205
 % [arXiv:0706.3543 [nucl-th]].
  %%CITATION = doi:10.1103/PhysRevC.76.065205;%%
  %41 citations counted in INSPIRE as of 04 Apr 2020

%\cite{Wang:2018fjm}
\bibitem{Wang:2018fjm}
  K.~L.~Wang, Q.~F.~L\"{u} and X.~H.~Zhong,
  Interpretation of the newly observed $\Sigma_b(6097)^{\pm}$ and $\Xi_b(6227)^-$ states as the $P$-wave bottom baryons,
  Phys.\ Rev.\ D {\bf 99}, 014011 (2019).
 % doi:10.1103/PhysRevD.99.014011
 % [arXiv:1810.02205 [hep-ph]].
  %%CITATION = doi:10.1103/PhysRevD.99.014011;%%
  %18 citations counted in INSPIRE as of 06 Apr 2020

%\cite{Wang:2017hej}
\bibitem{Wang:2017hej}
  K.~L.~Wang, L.~Y.~Xiao, X.~H.~Zhong and Q.~Zhao,
  Understanding the newly observed $\Omega_c$ states through their decays,
  Phys.\ Rev.\ D  95, 116010 (2017).
  %doi:10.1103/PhysRevD.95.116010
  %[arXiv:1703.09130 [hep-ph]].
  %%CITATION = doi:10.1103/PhysRevD.95.116010;%%
  %48 citations counted in INSPIRE as of 08 Jan 2020

%\cite{Xiao:2020oif}
\bibitem{Xiao:2020oif}
  L.~Y.~Xiao, K.~L.~Wang, M.~S.~Liu and X.~H.~Zhong,
  Possible interpretation of the newly observed $\Omega_b$ states,
  Eur.\ Phys.\ J.\ C {\bf 80}, 279 (2020).
 % doi:10.1140/epjc/s10052-020-7823-z
 % [arXiv:2001.05110 [hep-ph]].
  %%CITATION = doi:10.1140/epjc/s10052-020-7823-z;%%
  %3 citations counted in INSPIRE as of 06 Apr 2020

%\cite{Aaij:2020cex}
\bibitem{Aaij:2020cex}
  R.~Aaij {\it et al.} [LHCb Collaboration],
  First observation of excited $\Omega_b^-$ states,
  Phys.\ Rev.\ Lett.\  {\bf 124}, 082002 (2020).
%  doi:10.1103/PhysRevLett.124.082002
%  [arXiv:2001.00851 [hep-ex]].
  %%CITATION = doi:10.1103/PhysRevLett.124.082002;%%
  %10 citations counted in INSPIRE as of 06 Apr 2020

%%%%%%%%%%%%%%%%%%%%%%%%%%%%%%%%%%%%%%%%%%%%%%%%%%%



\end{thebibliography}
\end{document}